\documentclass[fleqn,12pt,twoside]{article}
\usepackage{espcrc1}

\usepackage{epsfig}
\usepackage[figuresright]{rotating}

\def\ltsim{\raise0.3ex\hbox{$<$\kern-0.75em\raise-1.1ex\hbox{$\sim$}}}
\def\gtsim{\raise0.3ex\hbox{$>$\kern-0.75em\raise-1.1ex\hbox{$\sim$}}}

\newcommand{\AmS}{{\protect\the\textfont2
  A\kern-.1667em\lower.5ex\hbox{M}\kern-.125emS}}

\title{Results of the Hydrodynamics Approach to Heavy Ion Collisions}

\author{Pasi Huovinen\address{School of Physics and Astronomy,
                              University of Minnesota\\
                              Minneapolis, MN 55455, USA}
}
       
\begin{document}

\maketitle

\begin{abstract}
Recent hydrodynamical calculations for Au+Au collisions at
$\sqrt{s_{NN}} = 130$ GeV energy are reviewed, and the initial
conditions of hydrodynamical evolution necessary to reproduce
experimental data are discussed.
\end{abstract}

\section{Introduction}
One of the first measurements at RHIC was the elliptic anisotropy of the
particles produced~\cite{Ackermann:2000}. It was found to be as large
as the hydrodynamical prediction~\cite{Peter00} and subsequent
hydrodynamical calculations achieved good fits to its centrality and
$p_T$ dependence~\cite{Kolb:2000,Huovinen:2001,Teaney:2001}.  In the
following I review how well hydrodynamical models can fit the data
published after those early papers and what constraints the data has
set on the initial state of hydrodynamical evolution.

\section{Comparison with the data}
   \label{data}

As an example of hydrodynamical results, I mostly quote the work of
Kolb and Heinz~\cite{Heinz:2002}. This is not meant to imply that other
calculations are of lesser importance, but rather to emphasize how much of
the data can be explained using one set of parameters.

	\subsection{Transverse momentum spectra}

\begin{figure}[tb]
 \begin{minipage}[t]{80mm}
   \begin{center}
     \epsfxsize 75mm  \epsfbox{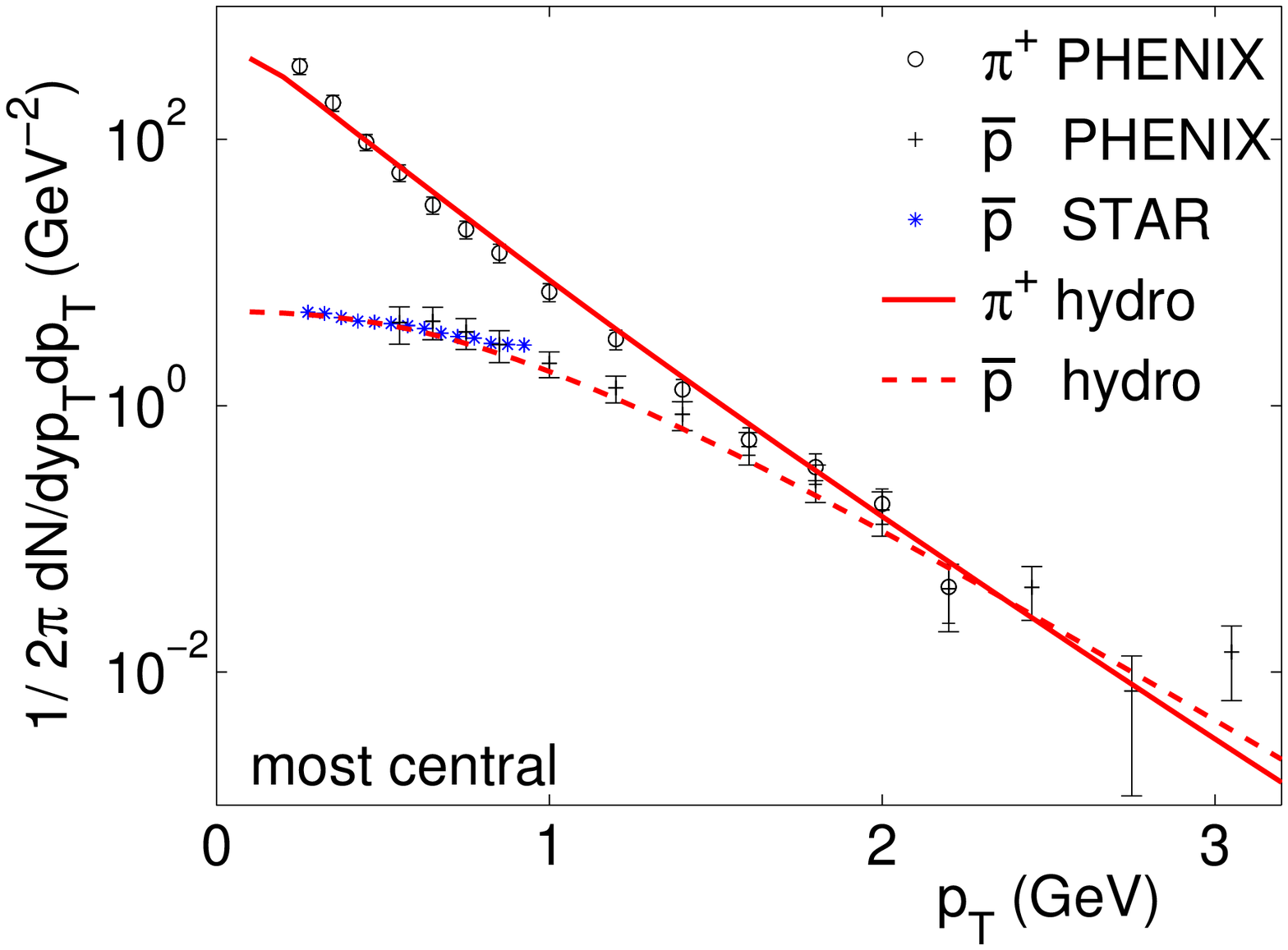}
   \end{center}
 \end{minipage}
\hfill
 \begin{minipage}[t]{80mm}
   \begin{center}
     \epsfxsize 75mm \epsfbox{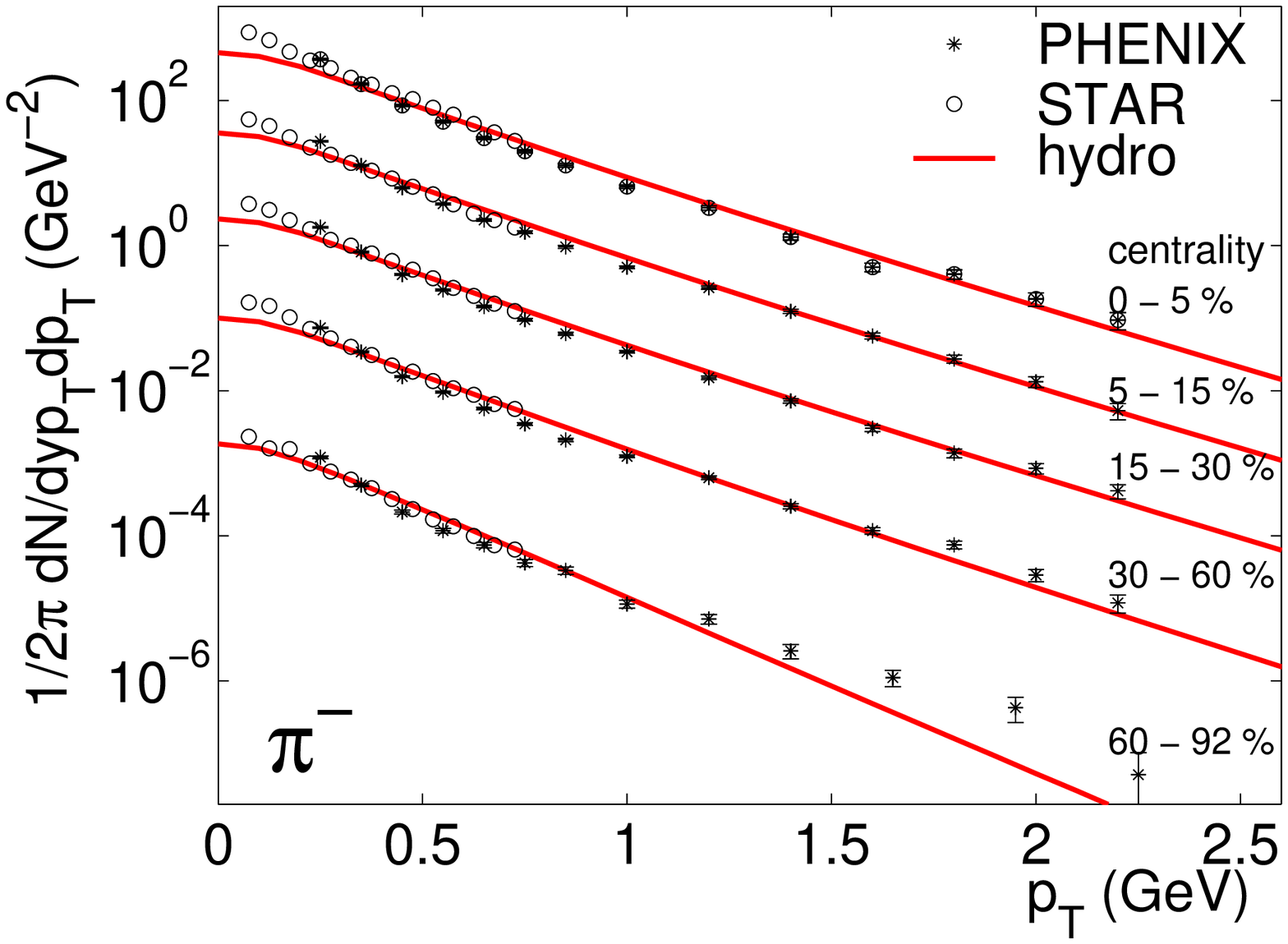}
   \end{center}
 \end{minipage}
\hfill
 \begin{minipage}[t]{80mm}
   \begin{center}
     \epsfxsize 75mm \epsfbox{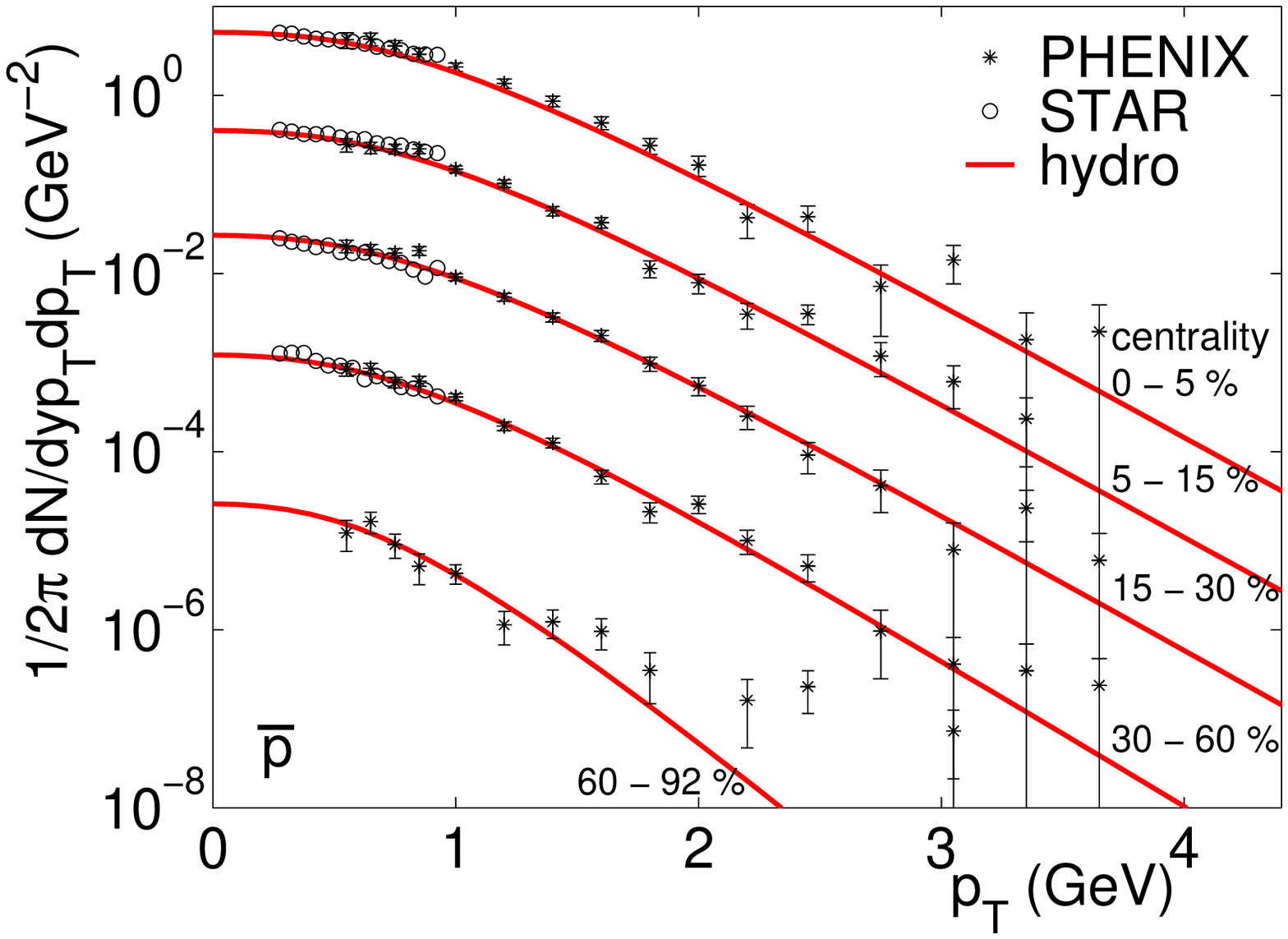}
   \end{center}
 \end{minipage}
\hfill
 \begin{minipage}[t]{80mm}
   \begin{center}
     \epsfxsize 75mm \epsfbox{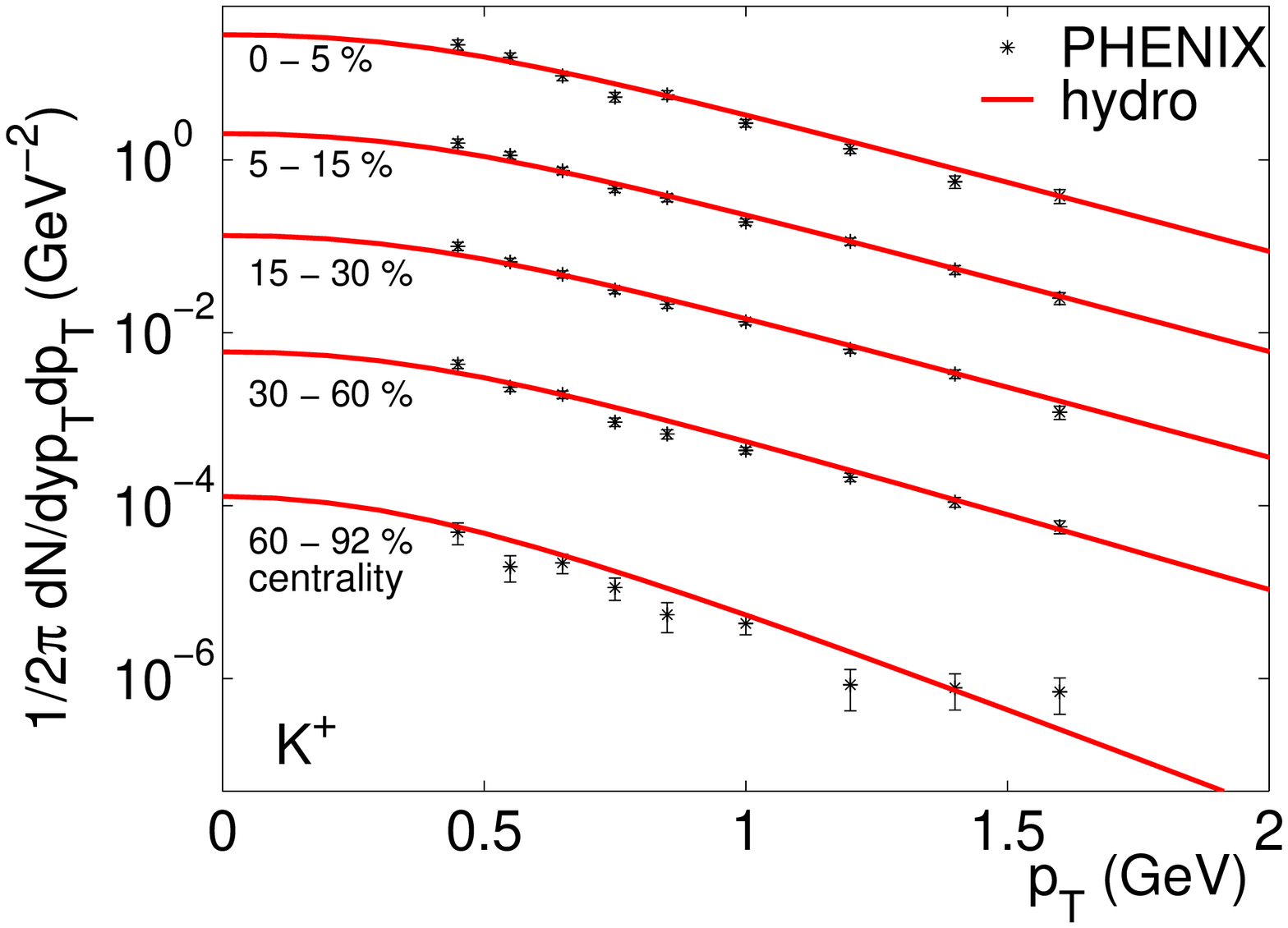}
   \end{center}
 \end{minipage}
\caption{Charged pion, antiproton and positive kaon spectra from central
         (upper left panel) and semi-central to peripheral (other panels) 
         collisions from~\cite{Heinz:2002}.
         The data were taken by the PHENIX~\cite{Velkovska:2001} and
         STAR~\cite{Adler:2001} collaborations.}
  \label{pts}
\end{figure}

The conventional method to initialize hydrodynamical calculations is
to fix the initial densities (energy or entropy and net baryon
density) to reproduce the observed particle abundancies and to choose
freeze-out temperature to reproduce the observed slopes of the
spectra. This approach was also applied in~\cite{Heinz:2002} where the
initial state was fixed to reproduce the charged particle multiplicity
as a function of centrality and the freeze-out temperature was chosen
to fit the central collision data. Therefore the good fit to pion and
antiproton spectra in the most central collisions shown in the upper
left panel of fig.~\ref{pts} is not surprising. On the other hand, the
slope of the spectra as a function of centrality are predictions and,
as can be seen, the agreement with data is impressive. For antiprotons
the prediction lies within the experimental errors up to $p_T=3$
GeV/$c$. Only for very peripheral collisions, $b > 10$ fm, do the data
show a significant excess of particles at $p_T>1.5$ GeV/$c$. In
hydrodynamical description the observed excess of antiprotons over
pions at $p_T \gtsim 2$ GeV/$c$ is simply a consequence of strong
transverse flow.

In this calculation local chemical equilibrium is assumed to hold
until kinetic freeze-out. Since the favoured kinetic freeze-out
temperature, $T_f = 130$ MeV, is much smaller that the chemical
freeze-out temperature, $T_{ch}=165$--175 MeV given by thermal
models~\cite{thermal}, it is not possible to obtain correct proton and
antiproton yields simultaneously. In~\cite{Heinz:2002} this was
circumvented by using pion yields at $T_f=130$ MeV as they were, but
scaling by hand all the other particle yields to their calculated
values at hadronization temperature $T_c=165$ MeV. The results
obtained in this way are very similar to those of~\cite{Teaney:2001}
where the hadronic phase is described by a
transport model and separate chemical and kinetic freeze-outs are
included. Thus one can consider the results of~\cite{Heinz:2002} to be
reasonable approximations.

	\subsection{Elliptic anisotropy}

Since the initial particle production is azimuthally symmetric,
azimuthal anisotropy of the final particle distributions is a signal
of rescatterings among the particles produced. More frequent
rescattering can be expected to lead to a larger anisotropy, and since
hydrodynamics assumes zero mean free path and thus an infinite
scattering rate, it provides an upper limit to observable
anisotropies. Anisotropy is quantified by measuring the harmonic
coefficients $v_n(y,p_T;b)$ of a Fourier expansion in $\phi_p$ of the
measured hadron spectrum
$dN/(dy\,p_T\,dp_T\,d\phi_p)$~\cite{Voloshin}. Anisotropy
characterized by a non-zero second coefficient, $v_2$, is called
elliptic anisotropy or elliptic flow~\cite{Ollitrault}.

\begin{figure}[t]
 \begin{minipage}[t]{80mm} \begin{center} \epsfxsize 75mm
 \epsfbox{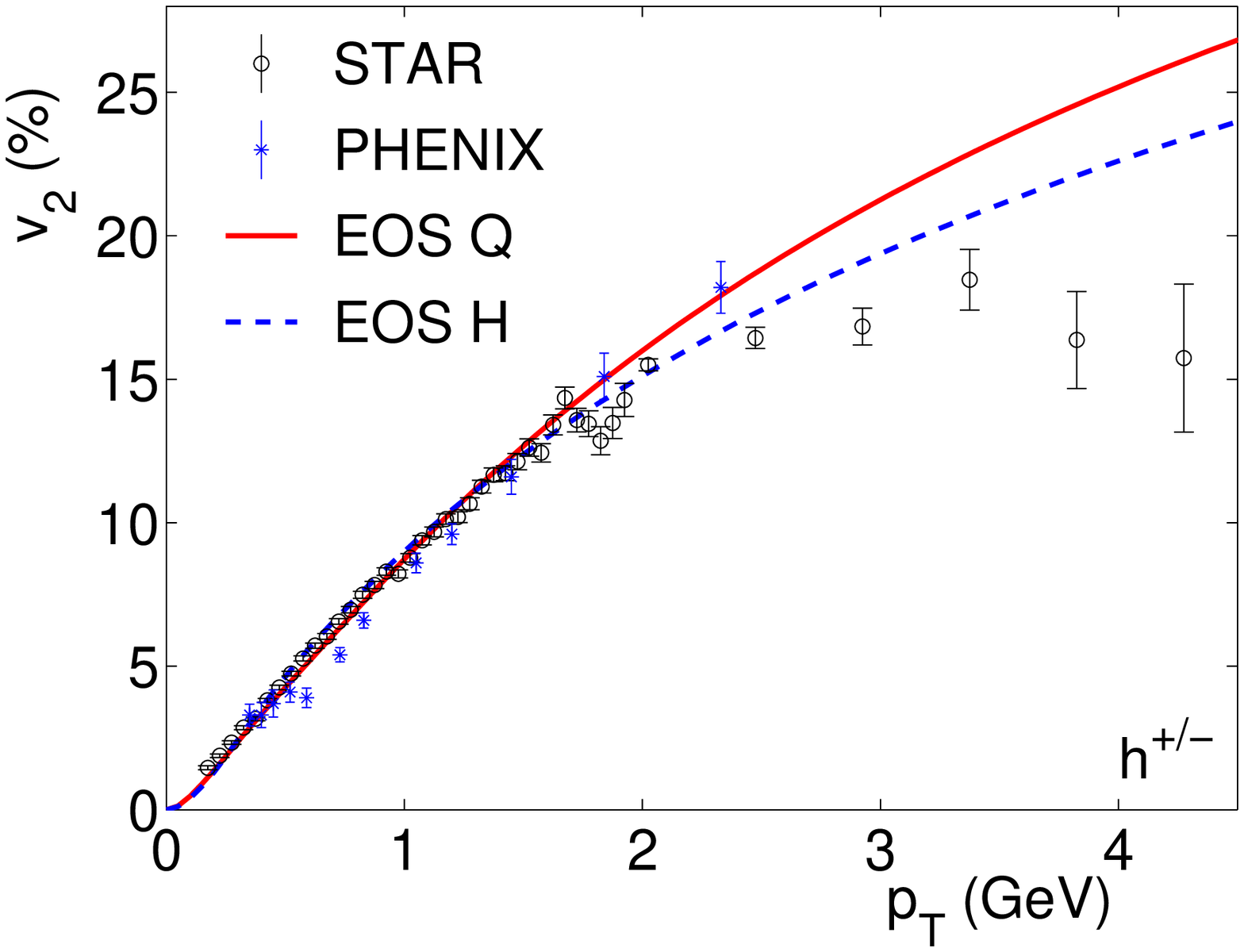} \end{center} \end{minipage}
\hfill
 \begin{minipage}[t]{80mm}
   \begin{center}
     \epsfxsize 75mm \epsfbox{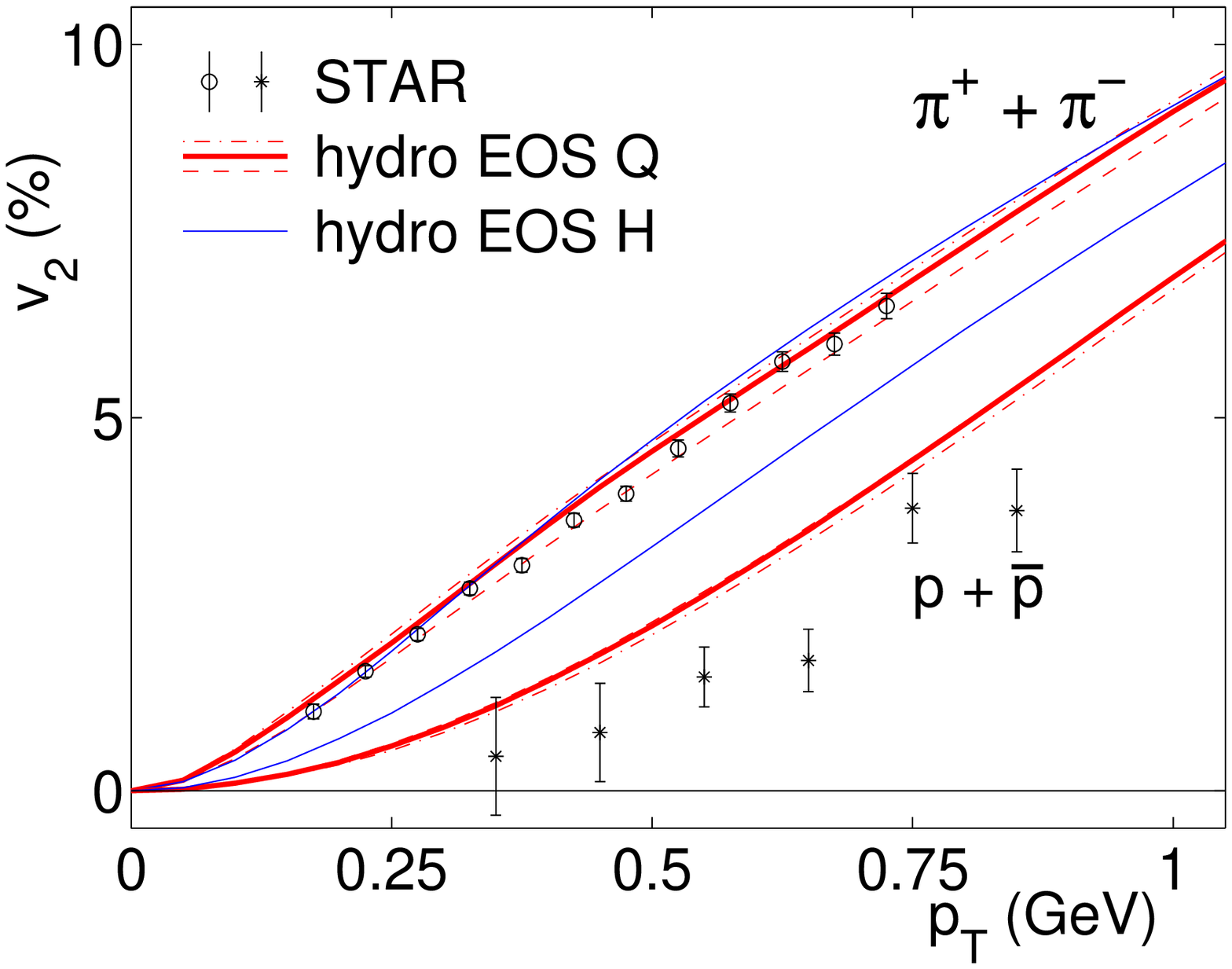}
   \end{center}
 \end{minipage}
\caption{The elliptic anisotropy coefficient $v_2(p_T)$
         of all charged particles (left) and identified pions and
         protons (right) at midrapidity in minimum bias collisions
         from \cite{Heinz:2002}. The data are from the
         STAR~\cite{Ackermann:2000,Star:v2pt,Snellings} and
         PHENIX~\cite{Lacey:2001} collaborations. The curves correspond
         to calculations using an equation of state with (Q) or without (H)
         a phase transition and (in the right panel) three different 
         freeze-out temperatures ($T_f = 128$ MeV (dash-dotted), 130 MeV
         (solid) and 134 MeV (dashed)).}
  \label{v2pt}
\end{figure}

The coefficient $v_2$ as a function of transverse momentum $p_T$ in
minimum bias collisions is shown in fig.~\ref{v2pt}. The calculation
was done using an equation of state with a first order phase
transition (EOS Q) and without a phase transition (EOS H).  The
observed anisotropy of charged hadrons shown in the left panel is seen
to reach the hydrodynamical values up to $p_T < 2$ GeV. Above $p_T=2$
GeV the hydrodynamical result keeps increasing monotonically whereas
the data saturates, indicating incomplete thermalization of the high
momentum particles. The pion and proton anisotropies depicted in the
right panel also show the hydrodynamically predicted mass dependence:
at low values of transverse momentum, the heavier the particle, the
smaller its $v_2$~\cite{Huovinen:2001}. Interestingly at this
conference the pion anisotropy was shown to deviate from
hydrodynamical predictions around $p_T=1.5$--2 GeV, whereas the proton
anisotropy reaches hydrodynamical values even at $p_T \sim 2.5$
GeV~\cite{phenix-here}. As can be seen the anisotropy of hadrons or
pions is quite insensitive to the phase transition, but the anisotropy
of protons shows a clear dependence: the curve calculated assuming a
phase transition is very close to the data, but an EoS without a phase
transition leads to too large a proton anisotropy at low $p_T$.

The fit to the proton data shown in fig.~\ref{v2pt} is worse than
achieved in earlier calculations~\cite{Huovinen:2001,Snellings}. This
is due to a different freeze-out temperature, which was $T_f=120$ MeV
in~\cite{Huovinen:2001}, but was chosen to be $T_f=130$ MeV
in~\cite{Heinz:2002} to achieve a better fit to the $p_T$ spectra. On
the other hand, the hydro + cascade approach of~\cite{Teaney:2001}
reproduces well both the $p_T$ spectra and differential anisotropy
$v_2(p_T)$ of pions and protons without such ambiguity. Whether this
points to a weakness in the ordinary Cooper-Frye description of
freeze-out or is a result of differences in the initialization of the
calculations remains to be seen.

\begin{figure}[tb]
\begin{minipage}[t]{75mm}
  \begin{center}
    \epsfxsize 65mm  \epsfbox{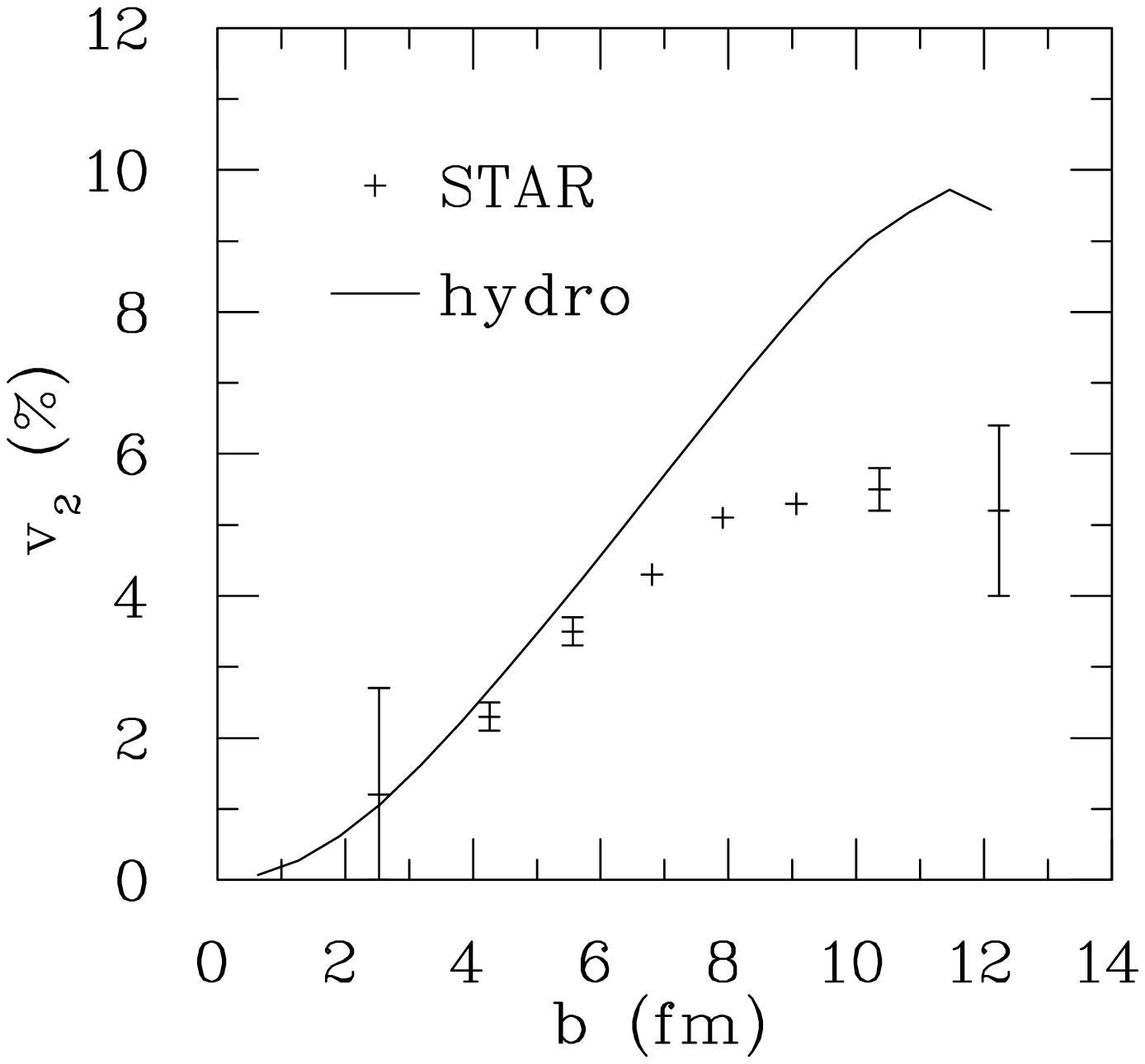}
  \end{center}
\caption{Centrality dependence of elliptic anisotropy of charged
         particles at midrapidity. The data are analysed using 4th order
         cumulant method by the STAR collaboration~\cite{STAR:4}.}
 \label{v2b}
\end{minipage}
\hfill
\begin{minipage}[t]{80mm}
  \begin{center}
    \epsfxsize 79mm \epsfysize 56mm \epsfbox{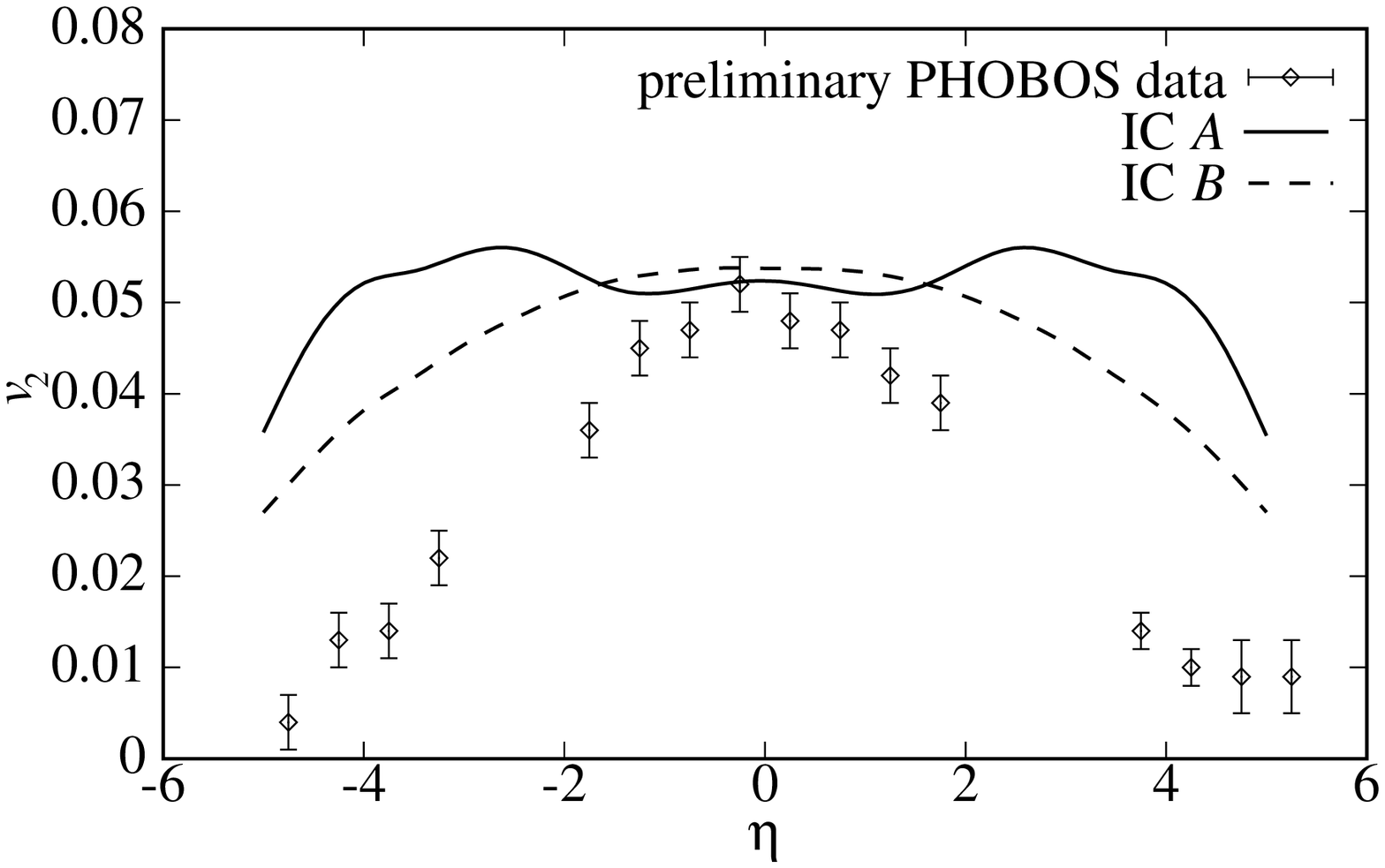}
  \end{center}
\caption{Pseudorapidity dependence of elliptic anisotropy of charged
         particles in minimum bias collisions from~\cite{Hirano:2001}.
         The preliminary data are from~\cite{Phobos-v2}. The curves
         correspond to two different parametrizations of the initial state.}
 \label{v2eta}
\end{minipage}
\end{figure}
\begin{figure}
  \begin{center}
    \epsfxsize 140mm \epsfbox{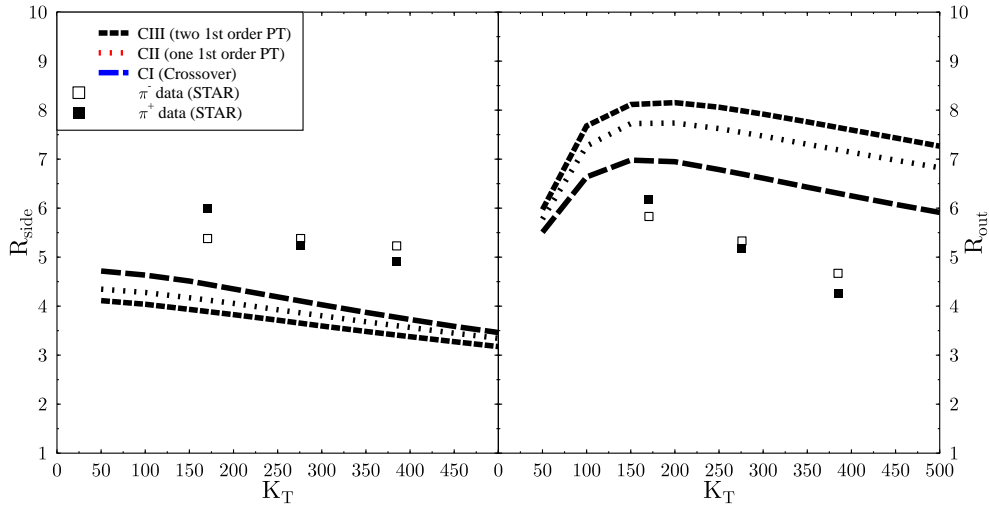}
  \end{center}
\caption{The HBT radii $R_{\rm side}$ and $R_{\rm out}$
         from~\cite{Zschiesche:2001}. The data are from~\cite{STAR-HBT}. 
         The curves correspond to equations of state with a first order
         phase transition but different latent heat (CIII and CII) or
         with a smooth crossover between the phases (CI).}
  \label{hbt-kuva}
\end{figure}

In fig.~\ref{v2b} the centrality dependence of the anisotropy
coefficient $v_2$ is shown. The calculation is similar to that
in~\cite{Heinz:2002} and the data is analysed by the STAR
collaboration~\cite{STAR:4} using the 4th order cumulant
method~\cite{Borghini}. The fit to the data is slightly worse than
shown before~\cite{Kolb:2000} since the 4th order method leads to
smaller values of $v_2$ than previously published~\cite{Ackermann:2000}.
The new values are, however, within the systematic errors of the older
analysis. The present data is below the hydrodynamical limit in all
but the most central collisions, but the discrepancy is significant
only for large impact parameters, $b>6$ fm.

In the calculations mentioned so far the expansion is assumed to be
boost invariant. If the assumption of boost invariance is relaxed and
expansion in all three dimensions is done numerically, the centrality
and $p_T$ dependence of the anisotropy at midrapidity is similar to
that in the boost invariant case~\cite{Hirano:2001}. When compared to
the excellent fit to data shown in fig.~\ref{v2pt}, the pseudorapidity
dependence of elliptic anisotropy shown in fig.~\ref{v2eta}
\cite{Hirano:2001} looks less satisfactory. The data reaches the
hydrodynamical value only around midrapidity. On the other hand, even
this result reproduces the data within a window of one to two units of
pseudorapidity. This region already contains most of the particles
produced. It is worth remembering that anisotropy in hydrodynamical
models depends strongly on the initial shape of the system. The
initialization used in~\cite{Hirano:2001} is relatively simple and a
more sophisticated initialization may lead to better fit to the
data. Therefore it is premature to conclude, based on this data and
calculation alone, that thermalization is reached only at midrapidity.

	\subsection{Two-particle interferometry}

The analysis of two-particle momentum correlations known as HBT
interferometry provides a method to study the space-time structure of
the emitting source~\cite{HBTreview}. It has been predicted that a
first order phase transition would lead to unusually large HBT
radii~\cite{Rischke:1996}. However, comparisons of
calculations~\cite{Heinz:2002,Soff,Zschiesche:2001,Morita:2002} with
data~\cite{STAR-HBT,Phenix-HBT} have lead to the so-called
\emph{HBT-puzzle}: All calculations give a ratio of HBT radii
$R_\mathrm{out}/R_\mathrm{side}$ larger than one, but the experimental
value is of order one. As shown in fig.~\ref{hbt-kuva}
\cite{Zschiesche:2001}, hydrodynamics usually leads to a too small
sideward radius $R_\mathrm{side}$ and to a too large outward radius
$R_\mathrm{out}$. This is usually interpreted to mean that the system
expands more and its lifetime is shorter than given by
hydrodynamics.

So far it looks like it is possible to fit either $R_{\rm side}$ or
$R_{\rm out}$, but not both simultaneously. A good fit to
$R_\mathrm{side}$ was achieved in~\cite{Soff} where the hadronic phase
was described using a hadronic cascade. This naturally leads to a
spatial freeze-out distribution with a larger average size in the
sideward direction and thus to larger $R_\mathrm{side}$. An acceptable
fit was achieved also in~\cite{Morita:2002} where the initial size of
the system was larger than in other calculations, but the outward
radius $R_\mathrm{out}$ is too large in both of these works.
In~\cite{Heinz:2002} it was shown that if freeze-out takes place
immediately after hadronization, $T_f=T_c=165$ MeV, $R_\mathrm{out}$
is reproduced, but $R_\mathrm{side}$ is even smaller than when the
freeze-out temperature is $T_f=130$ MeV. The effect of such a high
freeze-out temparture on single particle spectra or anisotropies was
not checked either.

One interesting detail shown in fig.~\ref{hbt-kuva} is that the radii
are closest to the data when an equation of state with a smooth
crossover is used. One can thus claim that HBT radii favour an
equation of state with crossover but, as argued above, proton $v_2$
results favour an equation of state with a first order phase
transition. The explanation to this seemingly contradictory
behaviour, as well as to the entire HBT-puzzle, is still unknown at
present. For the most recent developments see~\cite{here}.

\section{Initial conditions}
   \label{init}

It is well known that even if the hydrodynamical evolution is
constrained to reproduce the observed single particle spectra, there
still is considerable freedom in choosing the initial state. One
degree of freedom which complicates the comparison of the initial
state of different hydrodynamical calculations is the initial time,
$\tau_0$, of the calculation. Its effect is easily seen in the Bjorken
estimate~\cite{Bjorken,Adcox:2001} for the initial energy density,
\begin{equation}
  \epsilon_{Bj} = \frac{dE_T}{dy}\frac{1}{\tau_0\, \pi R^2},
\end{equation}
where a smaller initial time leads to a larger initial energy density
and vice versa. Thus a meaningful comparison of initial states from
different calculations requires scaling the initial densities to the
same initial time $\tau$. This is usually done by assuming one
dimensional Bjorken expansion with an ideal gas equation of state from
the initial time $\tau_0$ of a particular calculation to a chosen
common time $\tau$: $\epsilon = \epsilon_0 (\tau_0/\tau)^{4/3}$.

\begin{figure}[tb]
  \begin{center}
    \epsfxsize 75mm \epsfbox{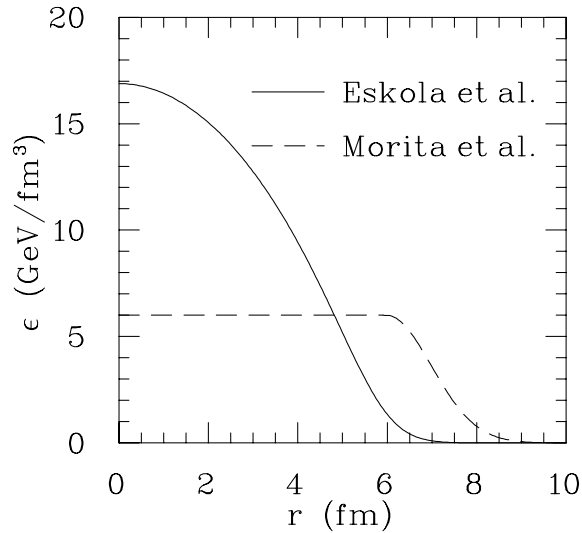}
  \end{center}
\caption{Initial energy density distributions in the transverse plane scaled
         to $\tau = 1$ fm/$c$ from~\cite{Morita:2002} (Morita {\em et al.}) 
         and~\cite{Eskola:2002} (Eskola {\em et al.}).}
   \label{initials}
\end{figure}

However, even if the difference in initial time is taken into account,
the final single particle spectra allow very different initial energy
distributions. As an example of this, initial energy distributions
used in recent calculations by Morita {\em et al.}~\cite{Morita:2002}
and Eskola {\em et al.}~\cite{Eskola:2002} are shown in
fig.~\ref{initials}. The former is a simple generalization of the one
dimensional Bjorken model to transversely expanding central
collisions: a flat energy density distribution with gaussian smearing
near the edges. The latter is based on a pQCD saturation
calculation~\cite{ekrt} of the transverse energy of minijets produced
in the primary collisions which leads to a very peaked distribution of
energy density. In this approach the net baryon density is also
obtained from pQCD calculation and --- after fixing the equation of
state and equating the initial time with the formation time,
$\tau_0=1/p_{\rm sat}$ --- the only free parameter is the freeze-out
temperature~\cite{errt}.

\begin{figure}
 \begin{minipage}[t]{85mm} 
   \begin{center}
      \epsfxsize 85mm  \epsfysize 74mm \epsfbox{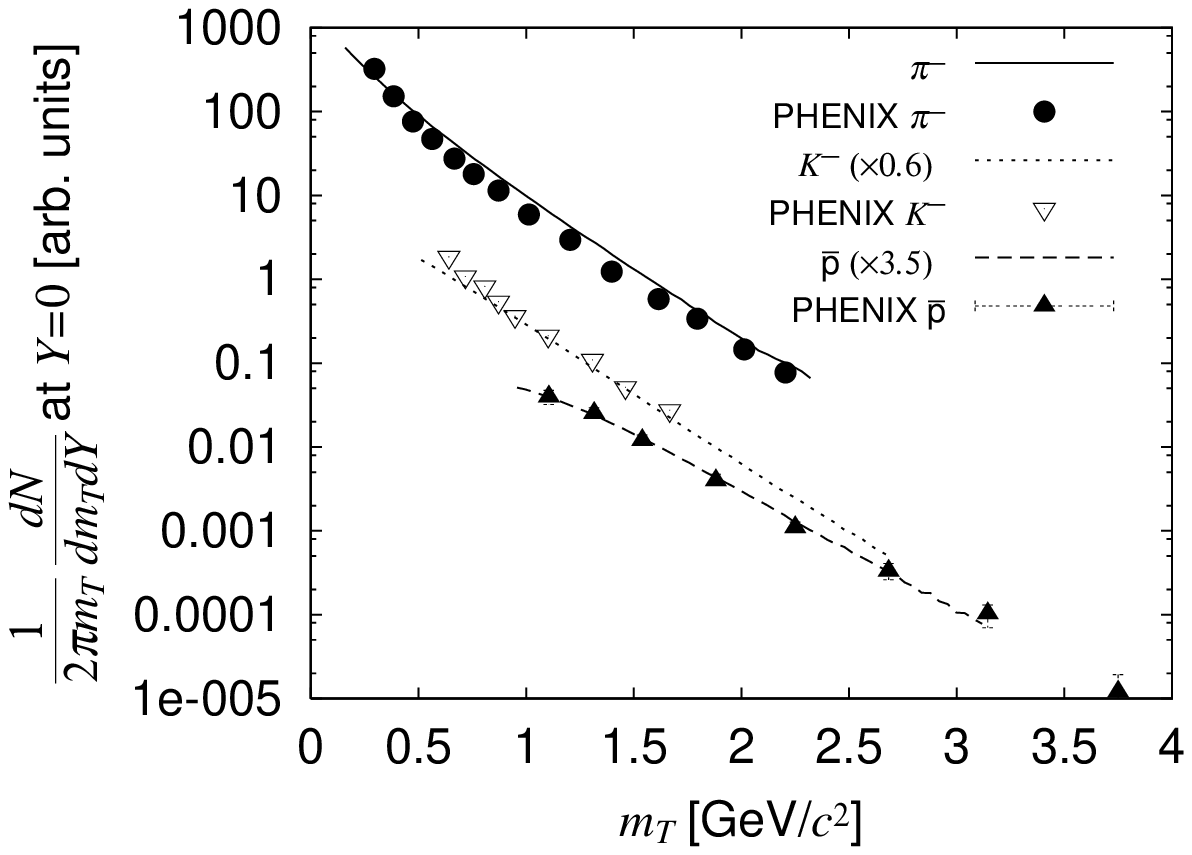}
   \end{center}
 \end{minipage}
\hfill
 \begin{minipage}[t]{75mm}
   \begin{center}
     \epsfxsize 70mm \epsfysize 72mm \epsfbox{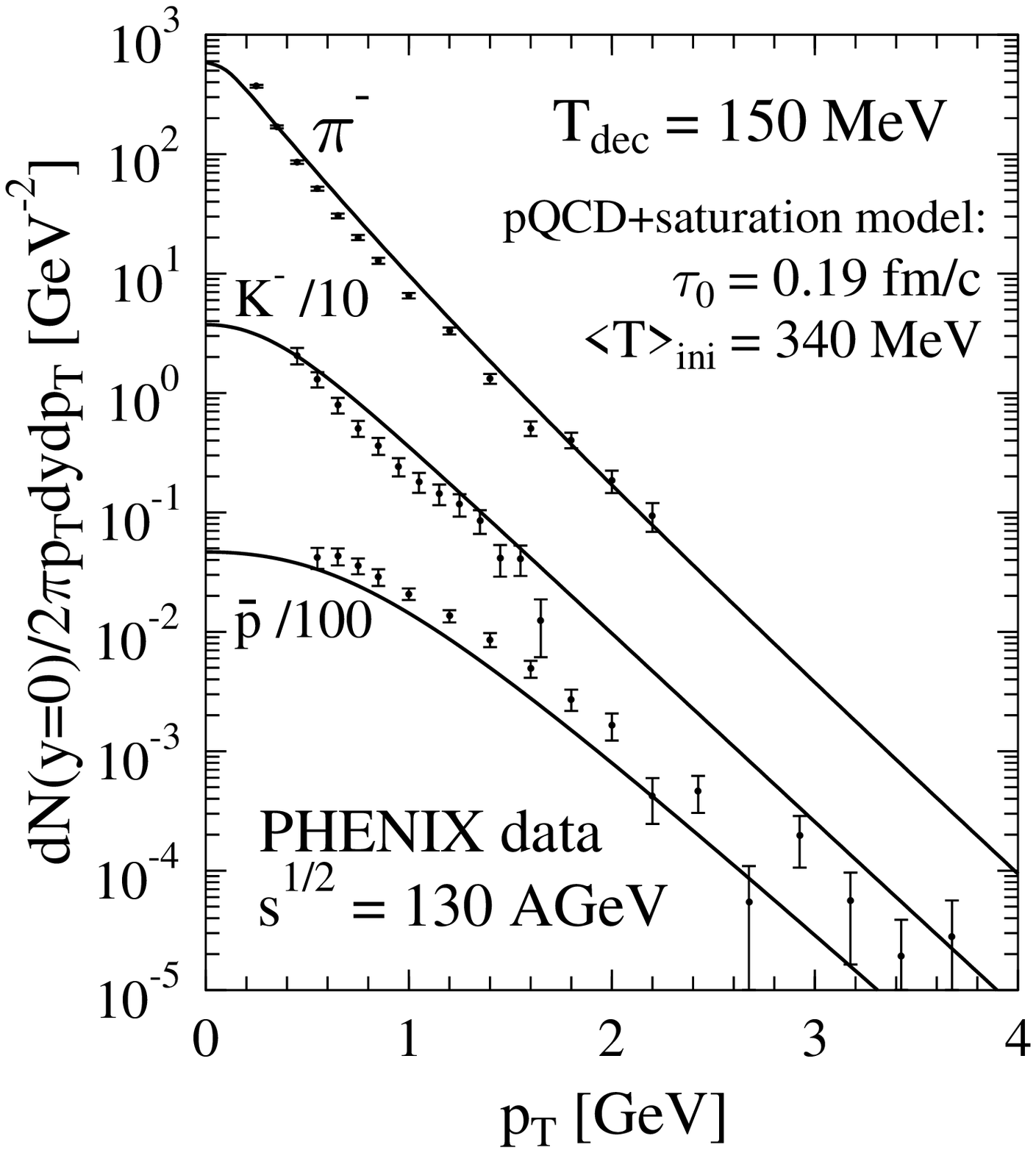}
   \end{center}
 \end{minipage}
\caption{The $p_T$ distributions of negative hadrons calculated by Morita
         {\em et al.}~\cite{Morita:2002}
         (left) and Eskola {\em et al.}~\cite{Eskola:2002} (right) compared
         to data by PHENIX collaboration~\cite{Velkovska:2001,Phenix-pt}.}
\label{pt-comp}
\end{figure}

Even though the initial states in these two calculations are very
different, the final single particle distributions shown in
fig.~\ref{pt-comp} are surprisingly similar and reproduce the slopes
equally well! The particle yields must again be treated with
caution. In both of these works chemical equilibrium is assumed to
hold until kinetic freeze-out. In~\cite{Morita:2002} the kinetic
freeze-out temperature $T_f=125$ MeV is well below the chemical
freeze-out temperatures given by thermal models. Thus the authors note
that their calculation reproduces the observed slopes of the spectra,
but to reach the yields, they must scale them by hand. On the other
hand, the authors of~\cite{Eskola:2002} employ a considerably high
freeze-out temperature, $T_f=150$ MeV, which allows them to
reproduce also the particle abundancies.

Unfortunately other observables are not helpful in differentiating these
two approaches either. Both of the calculations were done for central
collisions where the elliptic anisotropy is zero. The HBT radii are
sensitive to the build-up of flow and the lifetime of the system and
therefore to the initial distributions, but unfortunately HBT radii
were not calculated in~\cite{Eskola:2002}. As an estimate one can use
the HBT radii for a high freeze-out temperature shown
in~\cite{Heinz:2002} instead. The radii in~\cite{Heinz:2002}
and~\cite{Morita:2002} are indeed different, but neither of them
reproduces the data: the first leads to a correct $R_\mathrm{out}$ but
too small a $R_\mathrm{side}$, whereas the latter reproduces
$R_\mathrm{side}$ acceptably but $R_\mathrm{out}$ is too large.
Another observable which is sensitive to the maximum temperature
reached in heavy ion collisions is the spectrum of direct photons, but
there is no photon data available yet, nor was the yield calculated in
these papers.

Even if the initial energy density distribution is only weakly
constrained by the data, the initial energy per unit rapidity is almost
fixed. In Table~\ref{taulukko} the initial states of recent
hydrodynamical calculations are characterized by scaling
to $\tau=1$ fm/$c$ and calculating the average energy density in
the transverse plane. The peak values of the initial energy density have a
huge spread from 6 to 160 GeV/fm$^3$ in these works, but the average
values at 1 fm/$c$ have a much smaller spread from 3.9 to 6.5 GeV/fm$^3$.
This spread is partially due to the different values of the effective
transverse area, $\pi R^2$. If this is taken into account, the average
energy density in all works is 5--6 GeV/fm$^3$, very similar to the
the Bjorken estimate 4.6 GeV/fm$^3$~\cite{Adcox:2001}.

\begin{table}[tb]
 \caption{Average initial energy density in the transverse plane scaled to
          a time $\tau=1$ fm/$c$, initial time, and freeze-out temperature
          in recent hydrodynamical calculations.}
   \label{taulukko}
\newcommand{\m}{\hphantom{$-$}}
\newcommand{\cc}[1]{\multicolumn{1}{c}{#1}}
\renewcommand{\tabcolsep}{2pc} 
\renewcommand{\arraystretch}{1.2} 
 \begin{tabular}{@{}llll} \hline
		& $\langle \epsilon \rangle$(1 fm/$c$)
                                & $\tau_0$ [fm/$c$] 	 & $T_f$ [MeV] \\
 \hline
Kolb \& Heinz~\cite{Heinz:2002}	& 5.4 GeV/fm$^3$ &  0.6  & 130 \\
Hirano~\cite{Hirano:2001}	& 5.8 GeV/fm$^3$ &  0.6  & 137 \\
Teaney {\em et al.}~\cite{Teaney:2001} & $\sim 6$ GeV/fm$^3$ & 1.0 & N/A \\
Morita {\em et al.}~\cite{Morita:2002} & 3.9 GeV/fm$^3$ &  1.0  & 125 \\
Eskola {\em et al.}~\cite{Eskola:2002} & 6.5 GeV/fm$^3$ &  0.19 & 150--160 \\ 
  \hline
\end{tabular}
\end{table}

The initial times of hydrodynamic evolution employed in these
calculations have a relatively large spread from 0.2 to 1 fm/$c$.
However, in all calculations the initial time is shorter than the
perturbatively estimated thermalization time of a few
fm/$c$~\cite{Serreau}. The basic argument in favour of a short initial
time is based on elliptic anisotropy; the estimate by Kolb {\em et
al.}~\cite{Peter00} showed that if there is 1 or 2 fm/$c$ delay in
thermalization, the value of $v_2$ decreases by 10\% or 25 \%,
respectively, which leads to values of $v_2$ which are below the data.

There is no similar consensus about the freeze-out temperature at
RHIC. The single particle spectra in central collisions can be fitted
equally well (fig.~\ref{pt-comp}) if freeze-out takes place almost
immediately after hadronization at $T_f = 150$--160
MeV~\cite{Eskola:2002} or at $T_f=125$ MeV~\cite{Morita:2002}, roughly
at the same temperature as at the SPS. The elliptic anisotropy of
protons provides an additional constraint on the freeze-out
temperature. So far it seems to favour a relatively low freeze-out
temperature of $T_f \leq 130$ MeV, but whether it is possible to fit
the elliptic anisotropy data using same kind of initial state and
freeze-out temperature as in~\cite{Eskola:2002} remains to be
seen. While discussing freeze-out temperature at RHIC it is also worth
remembering that in the hybrid model of~\cite{Teaney:2001}, where the
plasma phase is described using hydrodynamics and the hadron phase using
hadronic cascade (RQMD), freeze-out temperature is not well defined.

\section{Summary}

Hydrodynamical models have been very successful in explaining the
single particle RHIC data at low $p_T$. The $p_T$ spectra and
anisotropies in central and semicentral collisions are well reproduced
for $p_T \le 1.5$ -- 2 GeV, and the $\bar{p}/\pi$ ratio at $p_T\sim 2$
GeV/$c$ has a simple explanation due to flow. Especially impressive
has been how hydrodynamics is able to create simultaneously an
elliptic anisotropy of negative hadrons which is large enough and an
anisotropy of protons which is small enough to fit the data. If one
considers solely this data the collision system behaves like a thermal
system.

However, the reproduction of the HBT radii has been unsuccessful so
far. It is unclear whether one should refine the final freeze-out
process, the hadronization process, or the initial state to reach an
acceptable description of the data. Especially puzzling is the fact
that the HBT radii seem to favour a relatively stiff equation of state
with a crossover phase transition, whereas the elliptic anisotropy of
protons favours a soft equation of state with a first order phase
transition.

The details of the initial state required to fit the data are not yet
completely fixed. The common features of all calculations are that the
collision system thermalizes rapidly --- initial time is $\tau_0\leq
1$ fm/$c$ --- and that the average initial energy density is well
above the critical energy density.

\section*{Acknowledgements}
This work was supported by the US Department of Energy grant
DE-FG02-87ER40328.

\end{document}